\documentclass[sigconf, nonacm %
]{acmart}

\usepackage{xurl}

\usepackage{paralist} %
\usepackage[inline]{enumitem}

\usepackage{fvextra}
\newcommand{\cd}[1]{\Verb|#1|}

\usepackage{fancyvrb}
\usepackage{csquotes}
\usepackage{paralist}
\usepackage{makecell}

\usepackage{multirow}

\newcommand{\rulesep}{\unskip\ \vrule\ }

\usepackage{pifont}

\usepackage{tikz}
\usepackage{pgfplots}
\usepackage{pst-plot}
\usepackage{pst-bar}
\usetikzlibrary{patterns,patterns.meta}
\pgfplotsset{
  every non boxed x axis/.append style={x axis line style=-},
  every non boxed y axis/.append style={y axis line style=-}
}
\makeatletter
\AddToHook{env/lstlisting/after}{\@endpefalse} %
\makeatother

\usepackage{pstricks-add}
\pgfplotsset{compat=1.18}
\usetikzlibrary{calc,arrows.meta,shapes.geometric,pgfplots.groupplots,scopes,backgrounds,patterns,patterns.meta}

\newcommand{\opsc}{\textit{Oopsie}}

\usepackage{fontawesome5}

\usepackage{enumitem}
\usepackage{subcaption} 

\usepackage[svgnames,dvipsnames]{xcolor}

\usepackage{siunitx}
\usepackage[nounderscore]{syntax}
\usepackage{algorithm}
\usepackage{algorithmicx}
\usepackage{algpseudocode}

\usepackage{listings}
\lstdefinestyle{bluejavacomments}{
morecomment=[l][\color{MidnightBlue}\bfseries]{//},
}
\lstdefinestyle{bluesqlcomments}{
morecomment=[l][\color{MidnightBlue}]{--},
}
\usepackage{fvextra}
\newcommand{\PS}{\texttt{Prepared\-Statement}}

\theoremstyle{plain}
\newtheorem{proposition}{Proposition}

\definecolor{cb1}{RGB}{127,201,127}
\definecolor{cb2}{RGB}{190,174,212}
\definecolor{cb3}{RGB}{253,192,134}
\definecolor{cb4}{RGB}{255,255,153}
\definecolor{cb5}{RGB}{56,108,176}
\definecolor{cb6}{RGB}{240,2,127}

\usepackage{lipsum}
\usepackage[
 disable
]{todonotes}

\begin{document}

\title{Static Type Checking for Database Access Code}

\author{Thomas James Kirz}
\orcid{0009-0001-1199-8801}
\affiliation{%
  \institution{University of Passau}
  \city{Passau}
  \country{Germany}
}
\email{thomas.kirz@uni-passau.de}

\author{Werner Dietl}
\affiliation{%
  \institution{University of Waterloo}
  \city{Waterloo}
  \country{Canada}
}
\email{wdietl@uwaterloo.ca}

\author{Mattias Ulbrich}
\affiliation{%
  \institution{Karlsruhe Institute of Technology}
  \city{Karlsruhe}
  \country{Germany}
}
\email{ulbrich@kit.edu}

\author{Stefanie Scherzinger}
\affiliation{%
  \institution{University of Passau}
  \city{Passau}
  \country{Germany}
}
\email{stefanie.scherzinger@uni-passau.de}

\begin{abstract}
JDBC remains a key technology for database access in Java applications. Since the database dictionary and the Java type system have distinct scopes, developers inevitably need to deal with bugs in SQL-to-Java type mappings. We propose an extension of the Java compiler, based on the established Checker Framework, which allows us to bridge this gap. Our approach verifies statically that the correct Java types are used when setting prepared statement parameters or when getting values from result sets. This allows us to lift a practically important class of runtime errors to compile time. Our approach is sound and, therefore, is guaranteed not to produce false negatives. Our prototype implementation also offers a degraded mode for type-checking legacy software, if developers are only interested in a subset of errors. Our experiments show that our approach detects a wide range of type mismatches in real-world application code and can indeed prevent errors which might otherwise surface as runtime errors. From the perspective of the developer, our approach is extremely lightweight: it processes the unmodified Java code, yet developers may add their own annotations. This allows us to perform type-checking even across method boundaries, whereas commercial developer tools are restricted to local checks. Finally, we show that we can type-check real-world JDBC software with reasonable overhead during compilation.
\end{abstract}

\maketitle

\section{Introduction}

Database applications form the backbone of modern information technology, and Java is a widely used language for their implementation.
Introduced nearly 30 years ago, JDBC~\cite{jdbc4.3} remains the go-to framework for 
 Java-to-SQL interaction. 
JDBC is part of the Java Standard Edition API and is actively maintained by Oracle.
Thus, 
JDBC-based applications enjoy long-term maintainability~\cite{7832897}.
This is particularly important in enterprise applications, which often remain in production for decades.

In the early 2000s, the first
object-relational mappers (ORMs) became popular in response to the verbosity 
of raw JDBC.
ORMs sit on top of JDBC and abstract SQL through object mappings. 
While the JDBC API has remained relatively unchanged over time, the ORM landscape has seen multiple generations, frameworks, and major redesigns.
In consequence, some developers outright dismiss ORMs in favor of %
plain vanilla JDBC. Developers who do choose to use ORMs oftentimes blend ORM code with raw JDBC code~\cite{10.1109/ICSM.2015.7332512,10.1007/978-3-319-39696-5-30,DBLP:journals/corr/DecanGM17}, since JDBC allows to formulate complex SQL queries that mappers may not express as easily or execute as efficiently~\cite{10.1145/2568225.2568259}.

Overall, JDBC has stood the test of time and remains crucial in database application development.
However, the apparent gap between the Java and SQL type systems makes JDBC code prone to certain kinds of bugs.
The examples below illustrate common pitfalls with database access code.

\begin{lstlisting}[
    style=bluejavacomments,
    escapechar=|,
    caption={Bugs in JDBC database access code.},
    label={lst:error-snippets},
    frame=single,
    numbers=left,
]
// 1. Typo in statement, table called "employee"
connection.prepareStatement("Select * from #employe#");|\label{ln:employe_typo}|

// 2. Wrong type of setter, 4th column is VARCHAR(100)
PreparedStatement ps = connection.prepareStatement(
  "INSERT INTO ROOMS VALUES (?,?,?,?)");
...
ps.set#Boolean#(4, room.isBooked());|\label{ln:setboolean_varchar}|

// 3. Wrong type of getter, result of count(*) is BIGINT
String sql = "select count(*) from USERS";  ...|\label{ln:select_count}|
int result = resultSet.get#Int#(1);|\label{ln:getint_count}|
\end{lstlisting}

\begin{example}\label{ex:intro}
Listing~\ref{lst:error-snippets}
shows snippets from open source Java projects.\footnote{Snippet~1 from \url{https://github.com/sayedabdul-aziz/JDBC-Course}, snippets~2 and~3 from
\url{https://github.com/iluwatar/java-design-patterns}. Snippets edited for readability.}  Line~\ref{ln:employe_typo} prepares a SQL statement. The typo in the table name causes
 a runtime exception.
Approaches to detecting malformed SQL statements already exist in published work~\cite{DBLP:conf/icse/GouldSD04a,10.1007/978-3-642-17164-2_10,10.1007/3-540-44898-5_1,1317486,DBLP:conf/icse/NagyC18,10.1007/978-3-319-39696-5-30}
 and are even implemented in commercial tools, e.g., ``IntelliJ IDEA Ultimate''~\cite{intellij_database_tools}.

However, the other problems above involve JDBC setter- and  getter-calls, and are not detectable by state-of-the-art tools.
In the second example, a Java Boolean is written into a column with SQL type \cd{VARCHAR(100)}.  While not \emph{recommended} by the JDBC specification, this mapping is nevertheless \emph{supported} by some JDBC drivers: these may then carry out an implicit type conversion without raising a runtime error, which is obviously  risky. This also constitutes a portability problem, as migrating to an new JDBC driver may also cause new runtime errors to materialize. 

The third example is similarly subtle. 
The query in line~\ref{ln:select_count} returns  
a BIGINT (i.e., 64-bit), yet the getter-call retrieving the value  only supports 32-bit Java integers. Again, this mapping is not recommended, but may be supported by some drivers. These may silently truncate the value, causing the problem to go unnoticed.
A call to \cd{getLong(1)} would have been the recommended choice. 
\end{example}

\emph{Practical relevance.}
The fact that that getter- and setter-calls in JDBC application code are not systematically being type-checked is a practically relevant problem:
(1)~Database access code is somewhat of a blind spot in software engineering.
Test coverage for database-related code is often poor~\cite{10.1016/j.is.2022.102105}, so type mismatches may materialize in production systems for the first time. 
(2)~In the software stack, the database schema often acts as a ``dependency magnet''~\cite{10.5555/1076577}: The strong coupling of the database schema with the  application code~\cite{10.1145/2491411.2491431} causes SQL queries to ``break'' when the database schema evolves~\cite{DBLP:journals/pvldb/CurinoMZ08}.
(3)~Legacy JDBC applications underpin large enterprise systems.
During decades of use, they will have to be migrated to new platforms, operating systems, database versions, and database drivers. This is typically done by teams other than the original developers, which amplifies the need for automated tools dedicated to analyze the code of database applications.
(4)~Via ``vibe coding'',  database application code is generated by AI coding assistants. %
Tools that can verify code correctness at compile time will play a crucial role in this new programming paradigm and are expected to become increasingly important.

\emph{Solution design.}
In this paper, we describe how to extend the Java type system with new SQL-specific types. This allows for static type-checks of getter/setter calls
and lifts a practically important class of JDBC-related runtime errors to compile time: 
    Code that passes type checking is \emph{guaranteed} not to throw runtime 
    exceptions related to type mismatches in JDBC getter/setter calls, nor will it carry out unsafe type conversions. We further guarantee the absence of getter/setter invocations with invalid parameter indices, or setters that use invalid column names. Existing approaches cannot do this. 

To be able to carry out these checks,
our approach \emph{automatically} infers SQL-related type information in Java code, as derived from the database schema. That is, application developers can type-check their code as is, without any changes. 

To demonstrate the immediate practicality of our solution,
we have implemented our static analysis in our prototype tool \opsc{}\footnote{Originally, the project was named the ``Optional Prepared Statement Checker (OPSC)'', but as the project outgrew this initial scope, we only kept the homophone \opsc{}.}.
\opsc{} has been designed to seamlessly integrate into the developer workflow: The minimal setup required is to simply configure access to the database data dictionary.
\opsc\ type-checking is then done during code compilation.

\paragraph{Contributions.}
This paper makes the following contributions.

\begin{itemize}[leftmargin=*]
   \item We present our type checker \opsc, an extension to the %
   Checker Framework~\cite{cf}.
   \opsc\ relies on anno\-tation-based type checking and statically detects problems with JDBC getters/setters that go unnoticed in existing analyzers. 
   We describe our approach for automatically inferring type annotations for JDBC database applications, given access to the database schema.

    \item Type-checking with \opsc\ is designed to be \emph{sound}. However, if  \opsc{} is applied to legacy code, it can be run in \emph{degraded mode}, in which unsupported, dialect-specific SQL features do not raise a compile time error, but are degraded to a warning. Subsequent getter/setter calls for such queries do not raise any further errors. This excludes the affected queries from the type guaranties, but still provides practical benefits when applied to existing code.%

    \item 
     Ours is the first approach for analyzing the SQL-related calls of JDBC code capable of \emph{modular, intra-procedural type checking}: 
    The type checker analyzes one method at a time, which allows it to scale well.    
    Our approach also supports intra-procedural static analysis where type-check\-ing getter/setter-calls requires an analysis across method boundaries, which is often not supported by other static analyses. Developers may \emph{optionally} add manual annotations to extend the reach of the tool.

    \item In our experiments, we type-check a diverse set of Java applications, including real-world software. Our experiments show that our approach can detect type mismatches more precisely than the commercial reference tool. 
    We further show that adding manual annotations is a very reasonable effort, enabling us to type-check up to 70\%--100\% of analyzable getter-calls in the Java projects under study.
    Finally, we show that runtime and memory overhead during code compilation are reasonable enough for \opsc\ to be used in practical software development.
\end{itemize}

\emph{Structure} This article is structured as follows. We review the preliminaries on JDBC and the Checker Framework in Section~\ref{sec:prelims}. We discuss related work in Section~\ref{sec:related}. In Section~\ref{sec:framework}, we present our  static analysis framework \opsc{}. Section~\ref{sec:experiments} presents our experiments with working open source Java code.
We discuss our insights in Section~\ref{sec:discussion}. We then conclude with an outlook on future work.

\section{Preliminaries}
\label{sec:prelims}

\subsection{JDBC Basics}

We introduce the JDBC API from the developer point-of-view and refer to the JDBC specification~\cite{jdbc4.3} for details.
In static code analysis, we target two Java interfaces that represent SQL statements, \cd{Statement} and \PS, and further  the JDBC interface \cd{ResultSet}, which represents the results of a SQL query.

A \cd{Statement}, as below, 
 takes a
SQL command string 
and sends it to the database for execution. This may produce a \cd{ResultSet}.

\begin{lstlisting}[
    style=bluejavacomments,
    escapechar=|,
    label={lst:jdbc_statement},
    frame=single,
    numbers=left,
]
Statement stmt = conn.createStatement();
String sql = "SELECT label FROM warehouse";
ResultSet rs = stmt.executeQuery(sql);

while (rs.next()) {
  int label = rs.getInt(1); // must match SQL type|\label{ln:getter_example}|
  System.out.println(label);
}
\end{lstlisting}

The \cd{ResultSet} is processed by iterating over its rows with \cd{next()}, and retrieving individual column values through type-appropriate getter methods.
Each getter requires either the column name or the 1-based column index as argument. Thus, the Java type of the getter-call in line~\ref{ln:getter_example} must be compatible with the 
SQL type of the attribute in the database schema.

The database system  can precompile SQL statements. In such prepared statements, the placeholders (\enquote{\cd?}) mark the positions of the parameters. Concrete values are later supplied via setter methods, such as \cd{setInt} or \cd{setDate}. The Java type is encoded in the method name and determines how the value is transmitted to the database.
Parameters are identified by their position (index) in the SQL statement.
Because the statement is precompiled, the \cd{execute} method does not take any input parameters.
This is illustrated by the code listing below. 

\begin{lstlisting}[
    style=bluejavacomments,
    escapechar=|,
    label={lst:jdbc_prepared_statement},
    frame=single,
    numbers=left,
]
String sql = "SELECT label FROM warehouse WHERE qty = ?";
PreparedStatement ps = conn.prepareStatement(sql);
ps.setInt(1, quantity); // Bind value of parameter '?'
ResultSet rs = ps.executeQuery();
while (rs.next()) {
    System.out.println(rs.getInt("label"));
} 
\end{lstlisting}

When executed repeatedly, with different parameter bindings, prepared statements can improve application performance. 
In addition, they play a crucial in preventing SQL injection~\cite{10.5555/2408046}.

\emph{Type compatibility.}
\label{sec:recommended-supported}
The JDBC specification distinguishes \emph{recommended} and \emph{supported}  conversions between Java and SQL types in getters/setters. 
A recommended conversion is part of the portable core of JDBC, and all conforming JDBC drivers implement these conversions. A driver may also implement a supported conversion, but is not required to do so. For any other conversion, a JDBC driver is not expected to allow it.
Attempting such a conversion usually results in a runtime exception. 
Figure~\ref{fig:jdbc-mapping}
highlights selected recommended and supported conversions for getters. 
For example, to access a value of SQL-type CHAR, calling \cd{getString} is recommended (first row in the table), while \cd{getInt} is only supported.
Likewise, the JDBC specification declares recommended and supported conversions for setters.

Note that the actual behavior of supported conversions may vary not only across JDBC drivers but also across database systems.
This is because drivers often rely on the underlying DBMS for type coercion, and DBMSs differ in their native type systems and casting semantics.
By sticking to recommended conversions, developers  ensure portability independent of both the driver and the DBMS.

\begin{figure}[t]
\centering

\footnotesize
\begin{tabular}{lll}
\toprule
SQL Type & Recommended getter & Supported getter \\
\midrule
CHAR     & getString  & getInt, getLong, getBoolean \\
VARCHAR  & getString  & getInt, getLong, getBoolean \\
INTEGER  & getInt, getLong     & getString, getBoolean \\
BIGINT   & getLong    & getString, getInt, getBoolean \\
\bottomrule
\end{tabular}

\caption{JDBC~4.3 type conversions~\cite{jdbc4.3} for getters (excerpt).}
\label{fig:jdbc-mapping}

\end{figure}

\emph{JDBC Metadata.}
Using JDBC, developers can retrieve metadata about the parameters in a prepared statement, %
and also the result of a query at run time.
This provides information such as column names and types. JDBC itself does not infer the schema of the query result, this task is instead delegated to the DBMS. Consequently, different DBMS products may return different metadata for the same query.

\subsection{JDBC Problems and Pitfalls}

The examples below illustrate common JDBC pitfalls
that cannot be caught by the Java compiler.

\begin{example}[Malformed SQL strings]
A malformed SQL statement string inevitably causes a runtime error when the query is executed.
In the code snippet below, we illustrate the straightforward case of a SQL syntax error (highlighted in red).

\begin{lstlisting}
stmt.executeQuery("SELECT * #FORM# warehouse");  #// Typo 
\end{lstlisting}

Line~2 in Listing~\ref{lst:error-snippets} shows another malformed SQL statement. It is syntactically correct but uses an incorrect table identifier. 
Yet another example of a malformed SQL string is shown in Listing~\ref{lst:oreilly-nonparsable}, where the parameter arities in the prepared statement do not match.
Such bugs can already be detected by existing developer tools.
\end{example}

\begin{example}[Problems with getters] \label{ex:problem_getters}
Below, we attempt to access the third column of a query result and the column \cd{id}.
Since there are only two columns and none of them are called \cd{id}, a  runtime error is raised.
We attempt to access the value of the first column as Java \cd{int}, yet the column has SQL type VARCHAR. This is not JDBC-recommended, yet existing developer tools do not detect this.

\begin{lstlisting}[numbers=left]
ResultSet rs = stmt.executeQuery(
  "SELECT label, qty FROM warehouse");
rs.next();
rs.get#Int#(1);       #// wrong type for VARCHAR column#
rs.getString(#3#);    #// invalid column index#
rs.getString("#id#"); #// invalid column label#
\end{lstlisting}
\end{example}

\begin{example}[Problems with setters] \label{ex:problem_setters}
Problems with setters are similar. %
Below, the setter in line~3 uses a non-recommended type conversion. Again, existing developer tools cannot detect this.
The setter in line~4 uses an invalid index, causing a runtime error. IDEs like IntelliJ IDEA Ultimate can detect mismatches between the number of placeholders in the query string, and the number of parameters set. This flags the issue in line~4.

\begin{lstlisting}[numbers=left]
PreparedStatement ps = conn.prepareStatement(
  "SELECT label FROM warehouse WHERE qty > ?");
ps.set#String#(1, "5");   #// wrong type for integer column#
ps.setString(#2#, "abc"); #// invalid parameter index#
\end{lstlisting}
\end{example}

\emph{Matters of scope.}
In the examples shown so far, the dynamic or prepared statement is always declared in the same method scope as the corresponding getter or setter calls. 
This is highly convenient for static code analysis, and even  considered best practice (and is actively encouraged by the Java \emph{try-with-resources} construct used to prevent resource leakage~\cite{bloch2008effective}).

However, database access code in real-world software frequently crosses method boundaries, yet tools like IntelliJ cannot perform code analysis under these conditions.

\subsection{Checker Framework}
\label{sec:prelims_checker}

The Checker Framework~\cite{cf,cfUses} is a powerful open-source framework for Java. 
It is actively used in real-world software development by major companies in industry and by research teams in academia.
It allows the implementation of pluggable type systems~\cite{Bracha04pluggabletype}: Extended type checkers plug into the normal Java compilation process using the annotation processing mechanism provided by the compiler infrastructure. This can be used to statically enforce properties like null-pointer exception freedom and correct usage of String interning~\cite{cfUses}.
The SQL Quotes Checker can analyze database applications to detect unescaped single quotes in SQL statements~\cite{CheckerFrameworkManual2022}, a vulnerability for SQL injection.
The Checker Framework further allows %
extended static analyses, such as inferring units of measurement~\cite{cfPUnit} or the combination of type systems with deductive verification~\cite{cfPropTypes} that go beyond the scope of typical type systems. 

A type system (a \enquote{checker}) for the Checker Framework consists of four main components which we have adapted and realized for \opsc{} in Section~\ref{sec:ann-based-type-chk}:
\begin{enumerate*}
\item the definition of the type qualifiers
  and the type hierarchy:
  this encodes the facts that are tracked about the program and their relationships;
\item the type introduction rules that determine the   qualified types for all source and bytecode elements;
\item optionally, the rules for enhanced flow-sensitive type refinement, which determines more specific types depending on the control flow of the program; 
and
\item the type rules that enforce correct behavior based on refined types, by traversing the Abstract Syntax Tree (AST) of the program.
\end{enumerate*}

A Java project can make use of multiple type systems using the Checker Framework, and information obtained from one type system can be used within another type system.

\definecolor{lightviolet}{rgb}{0.9,0.8,1.0}

\emph{Type annotations.}
\label{sec:value-checker}
The type qualifiers of a type system are expressed using Java annotations (denoted by an \cd{@} sign).
These annotations can be parametrized using \emph{elements}.
For example, the built-in \emph{Constant Value Checker}, which determines the value of a variable (if possible in static analysis), defines an annotation \cd{@IntVal} with an array element that tracks the value(s) that an integer variable may have.
Java code can be annotated manually, as in line~1 in the example below, or inferred automatically by the checker.

\begin{lstlisting}[
    style=bluejavacomments,
    escapechar=|,
    frame=single,
    numbers=left
]
x = 3; // OK (legal assignment)
\end{lstlisting}

If an input program cannot be typed according to the typing rules as implemented by the type system,
the checker plugin raises a compilation error.
In the example below, the inferred type of the right-hand side of the assignment in line~2 does not match the manually annotated variable type (line~1). This constitutes an illegal assignment. The program cannot be typed, and an error is raised.

\begin{lstlisting}[
    frame=single,
    numbers=left
]
x = 5; %
#// Error: incompatible types in assignment.
\end{lstlisting}

The Constant Value Checker also introduces a \cd{@StringVal} annotation to statically determine \cd{String} values. Our implementation
\opsc{}, to be introduced later, uses the \cd{@IntVal} annotation to determine the column arguments of setter and getter calls, as well as \cd{@StringVal} to extract SQL statement strings from Java code.

\emph{Modularity.}
Explicit type annotations are particularly useful in method declarations to achieve modularity---each method is type checked in isolation from all other code, only relying on local type information or information from other method declarations. 
This makes the type checks modular and intraprocedural, in contrast to other static analysis approaches that are global, require all source code, and perform interprocedural analyses.

\section{Related Work}
\label{sec:related}

\emph{Typing across software layers.}
The challenge of reconciling two isolated type systems, one in the application layer and one in the database layer, has been widely recognized. 
Frameworks such as Microsoft's LINQ~\cite{10.1145/1142473.1142552} demonstrate that strong type safety can be achieved through tight integration of programming and query languages.
Very recently, a new proposal has been made to radically re-design query languages for a more seamless experience in application development (e.g.,~\cite{DBLP:journals/corr/abs-2504-12953}).
While these approaches
share our overall objective, they require developers to adopt new languages and paradigms, and ultimately, to rewrite existing database applications.
In contrast, we target an established and commercially relevant market: 
database applications written in Java with JDBC, which includes a large number of legacy applications. %

\emph{Static analysis for JDBC applications.} A substantial body of work has addressed the static analysis of SQL queries embedded in JDBC code. This requires access to both the application code and the database schema, typically to detect syntactic and semantic errors in SQL statement strings (e.g.,~\cite{DBLP:conf/icse/GouldSD04a,10.1007/978-3-642-17164-2_10,10.1007/3-540-44898-5_1,1317486,DBLP:conf/icse/NagyC18,10.1007/978-3-319-39696-5-30})
like the first issue in Listing~\ref{lst:error-snippets}.  The challenge lies in analyzing dynamically constructed SQL statements, by combining automata-based techniques with control flow analysis.
This work is complementary to ours.

Once SQL statement strings have been extracted, a range of additional analyses becomes possible. These include the detection of ``SQL smells''~\cite{8090148,DBLP:conf/icse/NagyC18}, of inefficient queries~\cite{DBLP:conf/icse/NagyC18}, safeguarding applications against breaking schema changes~\cite{7589806}, and most prominently, detecting vulnerabilities related to SQL injection (e.g.,~\cite{10.5555/2408046,DBLP:journals/infsof/ThomasWX09}). Notably, the Checker Framework used in our work also provides a plugin to detect SQL injection attacks.

\emph{Commercial tools.}
Professional developer tools also provide support for JDBC development. For instance, the Java IDE ``IntelliJ IDEA Ultimate''~\cite{intellij_database_tools} offers syntax highlighting and schema-aware autocompletion.
In the context of our work, IntelliJ can detect obvious arity mismatches in JDBC prepared statements, as shown in Listing~\ref{lst:oreilly-nonparsable}, as well as obvious index mismatches in JDBC setter calls. This is achieved by comparing the number of placeholders in prepared statements with the indices used in setter access functions.
However, such tools are generally limited to very simple Java constructs, and struggle with non-trivial control flow. %

\begin{lstlisting}[%
caption={Mismatched parameter arities (found in~\cite{reese2000database})},label={lst:oreilly-nonparsable}]
INSERT INTO CUSTOMER (CUSTOMER_ID, FIRST_NAME, LAST_NAME, 
SOCIAL_SECURITY, CRT_CLASS, LUID, LUTS) 
VALUES (?, ?, ?, ?, ?, ?)
\end{lstlisting}

\emph{Summary.}
Existing approaches do not detect type mismatches in JDBC getter and setter calls, like the second and third examples in Listing~\ref{lst:error-snippets}, which our approach handles. While type checking JDBC getter and setter calls has been identified in prior work as a direction for future research~\cite{1317486},
we are not aware of any academic or commercial solution to this problem.
While IntelliJ IDEA only detects certain index mismatches for setters, our approach detects index mismatches also for getters, as well as type mismatches.

\emph{Own previous work.}
A very early version of our prototype implementation \opsc{} was presented at the BTW'25 student track program~\cite{DBLP:conf/btw/Kirz25}. 
The preliminary experiments presented do not involve any  real-world software and do not explore manual annotations.

\section{Static Analysis Framework}
\label{sec:framework}

We first provide an overview of the system architecture of our prototype implementation \opsc, and then present its internals.

\subsection{System Architecture}

\begin{figure}[tb]
    \centering
    \includegraphics[width=\columnwidth]{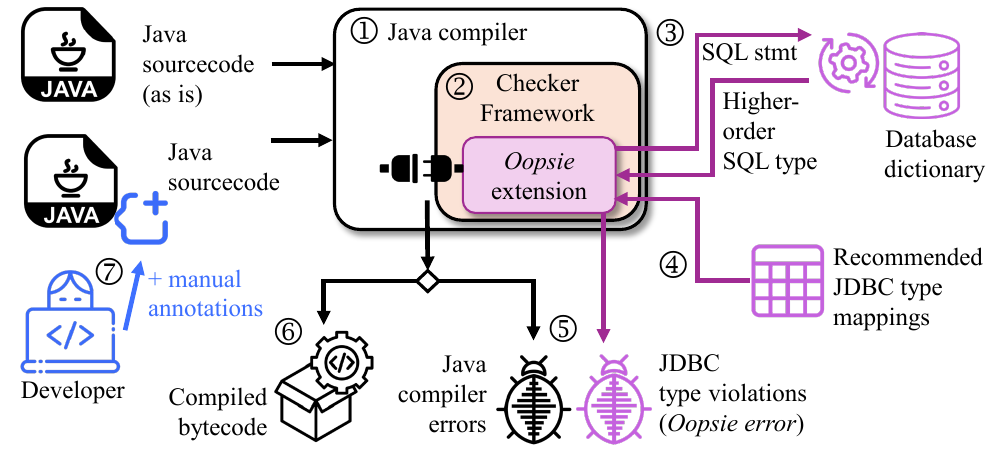}
    \caption{\opsc{} System architecture.}
    
    \label{fig:architecture}
\end{figure}

Figure~\ref{fig:architecture} shows the \opsc{} system architecture.  \opsc{} is designed to be part of the Java build process~(\textcircled{\scriptsize 1}). 
The heart of \opsc{} is an extension to the Checker Frame\-work~(\textcircled{\scriptsize 2}), which works as a plugin to the Java compiler and extends its capabilities. This extension parses SQL statement strings from JDBC \cd{Statement}s and \cd{PreparedStatement}s in the sources, and connects to a library with access to the database dictionary. At this point, \opsc\ can recognized malformed SQL statement strings.

This library~(\textcircled{\scriptsize 3}) further derives the expected SQL types of all parameters in parameterized statements, as well as the column names and SQL types of the attributes in the query result. \opsc{} captures this type information in automatically generated code annotations.
For example, this allows \opsc{} to resolve ``\verb!SELECT *!''-queries into higher-order SQL types, which list the order and attributes (with name and type) in the result set.

The \opsc{} extension has access to the JDBC-recom\-men\-ded type mappings~(\textcircled{\scriptsize 4}). Given the code annotations, the Java compiler can type-check the Java types in JDBC getter- and setter-calls against the expected SQL types. Thanks to the Checker Framework, this typing information is automatically propagated along the Java control flow (i.e., conditionals and loops).
If the type checker encounters a type mismatch,
the \opsc{} extension reports an \opsc{} \emph{error}~(\textcircled{\scriptsize 5}). These errors are generated in addition to the  Java compiler messages.
If no errors are detected, the Java code is compiled to bytecode (\textcircled{\scriptsize 6}), yielding the same compilation result as without \opsc.

In most scenarios, the approach does not require any changes to the code, because the type annotations needed by the tool chain can be \emph{inferred} fully automatically. To enable 
modular type checking across method boundaries, e.g., when a JDBC prepared statement is passed to or returned from a method, additional type annotations are required. They can be manually declared by the developers~(\textcircled{\scriptsize 7}).

\subsection{Guarantees and Limitations}

By construction, values computed only at run time are generally inaccessible to static analysis; accordingly, \opsc{} requires SQL strings and column references  to be known at analysis time. The approach supports literal values and, via the Checker Framework Constant Value Checker, can go beyond literal values and support some statically inferable non-literal values (see Section~\ref{sec:value-checker}). If the checker cannot determine the query string or the getter/setter argument, or if the query is malformed, \opsc{} reports an error.

\opsc's objective is to prevent certain categories of JDBC exceptions. When a Java developer writes a program for which the \opsc{} checker does not raise a compiler error, they can be sure that these exceptions will never occur when running the program:

\begin{proposition}\label{prop:1}%
If the \opsc{} type checker does not raise a type error for a program $P$, then \cd{SQLException}s will never be thrown at run time when executing $P$ as a result of \ldots
  \begin{enumerate}
  \item SQL syntax errors and schema mismatches when executing a \cd{Statement} or \cd{PreparedStatement},
  \item invoking setters with invalid parameter indices,
  \item invoking getters with invalid column names or indices, and
  \item invoking getters/setters with %
  types not recommended by the JDBC specification. %
  \end{enumerate}
\end{proposition}

Thus, using \opsc{} eliminates these categories of JDBC-related run-time failures. \opsc{} is \emph{sound} in the sense that it does not produce false negatives, although it may produce false positives (e.g., for dynamically computed queries). \opsc{} does not prevent all SQLExceptions: failures such as connection errors or uninitialised connections remain possible and must be handled by the application. At present, \opsc{} does not enforce that every placeholder of a prepared statement has been assigned a value; this is a very feasible extension that we plan to add in a future version.

As discussed in Section~\ref{sec:related}, other tools also address the first feature in Proposition~\ref{prop:1}, yet features (2)--(4) are unique to \opsc. %

\emph{Degraded mode.} \label{sec:degraded}
When a Java project is developed in a green-field approach and \opsc{} is part of the build chain from the start, \opsc{} reports compilation errors
promptly; developers can then modify the code such that it can be handled by the type checker and thus
 benefit from the guarantees in Proposition~\ref{prop:1}.

For retrofitting existing code, such strict enforcement may be impractical.  In many legacy code bases, not all calls will meet the
 restrictions required by the checker, and extensive refactoring solely to satisfy the type checker  is often infeasible. To accommodate this, \opsc{} provides a \emph{degraded mode} in which JDBC \cd{Statement}s and \cd{PreparedStatement}s whose SQL statement strings cannot be analyzed at compile time remain unchecked
 together with their associated getter/setter calls. This mode sacrifices global soundness but the type system offers \emph{local guarantees} that the checker preserves and verifies:

\begin{proposition}\label{prop:2}
  If the \opsc{} type checker does not raise an error at compile time
  for a program $P$ in degraded mode, then the exceptions mentioned in Prop.~\ref{prop:1} will never be
  raised \emph{by the statements covered by the type checker} at run time when executing $P$.
\end{proposition}

The type checking analysis is performed over the entire program modularly, proceeding method by method. This also allows our approach to scale.
However, not every type system that ensures run time guarantees enjoys this locality property. Im some type systems, a type constraint that is violated locally may manifest as a runtime error in a different region of the program later.

Our experiments in Section~\ref{sec:experiments} examine both scenarios: small, self-contained examples that illustrate the full benefits of \opsc, and  studies that demonstrate the benefits of applying \opsc{} to existing code bases, where \opsc{} operates in degraded mode yet still yields measurable benefits.

\subsection{Annotation-based Type-Checking for SQL}
\label{sec:ann-based-type-chk}

We now describe type-checking of JDBC getter/setter-calls within the  \opsc{} extension to the Checker Framework.
Our description follows the structure introduced in Section~\ref{sec:prelims_checker}:
We describe our type qualifiers and type hierarchy in Section~\ref{sec:opsc-annotations}.
Usually, \opsc{}  can automatically assign the \cd{@Sql} annotation to source code when certain method calls are performed; these type introduction rules are detailed in Section~\ref{sec:inferred-annos}.
Flow-sensitive refinement of the types is then discussed in Section~\ref{sec:type-refinement}.
After annotations have been inferred, setter and getter methods are checked based on the information stored in the annotations, as outlined in Section~\ref{sec:verifying-calls}.
Section~\ref{sec:manual-annos} describes how, in some cases, manually annotating the application code allows \opsc{} to check additional statements.

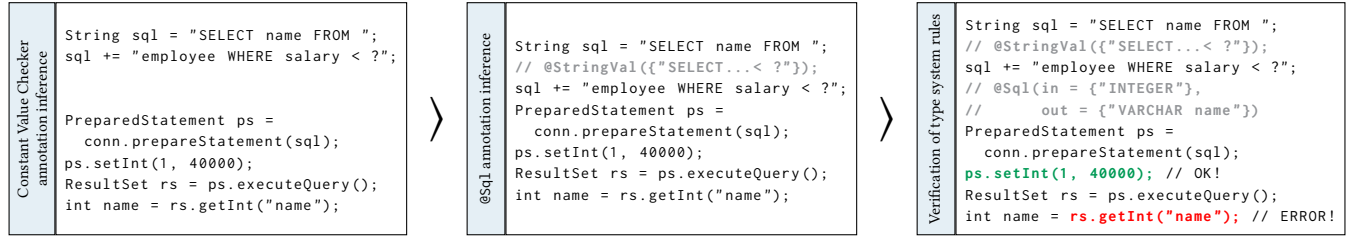
\begin{figure*}
    \centering
    \begin{tikzpicture}[
    nodes={align=center,font=\scriptsize},
    stepbox/.style={draw,fill=MidnightBlue!7,minimum width={3.07cm},minimum height=0.5cm,rotate=90,anchor=south east},
    codebox/.style={draw,minimum height=3.07cm},
]

\node[codebox] (code-value) {
\begin{lstlisting}[basicstyle=\ttfamily\scriptsize,frame=off]
String sql = "SELECT name FROM ";
sql += "employee WHERE salary < ?";


PreparedStatement ps = 
  conn.prepareStatement(sql);
ps.setInt(1, 40000);
ResultSet rs = ps.executeQuery();
int name = rs.getInt("name");
\end{lstlisting}
} ;

\node[codebox,right=1.3cm of code-value] (code-sql1) {
\begin{lstlisting}[basicstyle=\ttfamily\scriptsize,frame=off]
String sql = "SELECT name FROM ";
$// @StringVal({"SELECT...< ?"});$
sql += "employee WHERE salary < ?";
PreparedStatement ps = 
  conn.prepareStatement(sql);
ps.setInt(1, 40000);
ResultSet rs = ps.executeQuery();
int name = rs.getInt("name");
\end{lstlisting}
} ;

\node[codebox,right=1.3cm of code-sql1] (code-verify) {
\begin{lstlisting}[basicstyle=\ttfamily\scriptsize,frame=off]
String sql = "SELECT name FROM ";
$// @StringVal({"SELECT...< ?"});$
sql += "employee WHERE salary < ?";
$// @Sql(in = {"INTEGER"},$
$//      out = {"VARCHAR name"})$
PreparedStatement ps = 
  conn.prepareStatement(sql);
&ps.setInt(1, 40000);& // OK!
ResultSet rs = ps.executeQuery();
int name = #rs.getInt("name");# // ERROR!
\end{lstlisting}
} ;

\node[stepbox] (step-value) at (code-value.north west) {Constant Value Checker\\annotation inference} ;
\node[stepbox] (step-sql1) at (code-sql1.north west) {\cd{@Sql} annotation inference} ;
\node[stepbox] (step-verify) at (code-verify.north west) {Verification of type system rules} ;

\node at ($(code-value.east)!.5!(step-sql1.north)$) {\Huge$\rangle$} ;
\node at ($(code-sql1.east)!.5!(step-verify.north)$) {\Huge$\rangle$} ;

\end{tikzpicture}
    \caption{Steps of inferring \opsc{} annotations and checking JDBC access code. The annotations inferred in each step are highlighted in {\color{MidnightBlue} blue} and setter/getter-calls identified as correct or buggy are highlighted in {\color{ForestGreen} green} and {\color{red} red}, respectively.}
    \label{fig:comic}
\end{figure*}

\subsubsection{\opsc{} Annotations}\label{sec:opsc-annotations}

At the heart of the \opsc{} type system is the annotation type \cd{@Sql}.
Its two elements, \cd{in} and \cd{out}, store the SQL types for parameters and the result of a SQL statement.\footnote{The \opsc-internal elements \cd{file}, \cd{line}, and \cd{column} are used to track statements.}

We provide an EBNF grammar describing a \cd{@Sql} annotation with \cd{in} and \cd{out} elements below.

\begin{grammar}
<sql> ::= `@Sql(' [ <in_list> `,' ]  <out_list> `)' ;

<in\_list> ::= `in = {' <in_type> \{ `,' <in_type> \} `}' ; 

<out\_list> ::= `out = {' <out_type> \{ `,' <out_type> \} `}' ; 

<in\_type> ::= `"' <sql_type> `"' ;

<out\_type> ::= `"' <sql_type> [ <identifier> ] `"' ;

<sql\_type> ::= `INTEGER' | `VARCHAR' | `TIMESTAMP' | ...  ;

<identifier> ::= A result column identifier, e.g., a column name.
\end{grammar}

\begin{example} \label{ex:anno-examples}
The code snippet %
below illustrates a prepared statement and a dynamic statement with annotations that satisfy the grammar. 
Note that only \cd{PreparedStatements} have parameters,
so the \synt{in\_list} need not be declared for \cd{Statement}s.
For query results, the \synt{out\_list} also states the column names.

\begin{lstlisting}[
    frame=single,
    numbers=left
]
PreparedStatement ps = conn.prepareStatement(
  "SELECT id, salary FROM employee where dob = ?");

Statement stmt = conn.createStatement();
ResultSet rs = stmt.executeQuery(
  "SELECT username, dob FROM employee");
\end{lstlisting}
\end{example}

A \cd{@Sql} annotation type $\mathbf A=\texttt{@Sql(in=in1,out=out1)}$ may be a subtype of a type $\mathbf B=\texttt{@Sql(in=in2,out=out2)}$, depending on their elements.
This is the case if
\begin{enumerate*}
\item \cd{out2} is a prefix of \cd{out1}, i.e. \cd{out2} has $n$ entries and the first $n$ entries of the two lists are equal, and
\item \cd{in1} and \cd{in2} are equal.
\end{enumerate*}
This allows an $\mathbf A$-value to be used where a $\mathbf B$-value is expected, as all legal getters for $\mathbf B$ are also legal for $\mathbf A$.
The \cd{in} lists must be equal to allow verifying that all parameters have been set in a future feature.

The Checker Framework requires the type hierarchy to be a bounded lattice; \opsc{} defines the \cd{@SqlBottom} and \cd{@SqlUnknown} types as \emph{bottom} and \emph{top} types, i.e.\ universal super- and subtypes at the respective ends of the type hierarchy.
To mark statements as unsupported and to avoid redundant error messages, the type \cd{@SqlUnsupported} is introduced. The annotations \texttt{@Creates\-Sql\-State\-ment} and \texttt{@RetrievesSql\-Re\-sult\-Set} (outside of the \cd{@Sql} hierarchy) mark methods which produce statement or result set objects according to a provided SQL string, as we will explain further on. This avoids hard-coding the set of these special methods.

\subsubsection{Inferring Annotations Automatically}\label{sec:inferred-annos}

\opsc{} automatically introduces \cd{@Sql} annotations to statements and result sets and can therefore automatically insert annotations like the two blue ones in Example~\ref{ex:anno-examples} during the type checking process.

This is done by overriding the \cd{visitMethodInvocation} me\-thod provided by the Checker Framework, which lets us annotate the abstract syntax tree of the application code.
Each method invocation found in the syntax tree is analyzed in sequence.
Because we only want to annotate objects created by methods known to produce JDBC statements and result sets, the logic is restricted to invocations of methods annotated with \cd{@CreatesSqlStatement} or \cd{@RetrievesSqlResultSet}.

We used Checker Framework mechanisms to provide {\tt @Creates\-Sql\-Statement} annotations.
\opsc{} detects all statements created by calls to the methods \texttt{execute\-Query}, \cd{executeUpdate}, \cd{execute} and \texttt{execute\-Large\-Update} of the Java interface \cd{java.sql.Statement}, as well as all prepared statements created by the \cd{prepareStatement} methods of \texttt{java.sql.Connection}.

For methods that are annotated as \cd{@CreatesSqlStatement},
\opsc{} uses the Constant Value Checker (introduced in Section~\ref{sec:prelims_checker}) to extract the raw SQL statement strings from the corresponding String arguments to these functions.

\begin{example}   
Figure~\ref{fig:comic} shows the steps of inferring \opsc\ annotations.
In the first step, the Constant Value Checker extracts a dynamically concatenated string. This string is captured by the generated annotation \cd{@StringVal} for further analysis.
\end{example}

Next, the result in/out types can be determined for the SQL statement, by accessing the data dictionary. Then, the statement objects are annotated with a \cd{@Sql} annotation accordingly.

\begin{example}   
Step~2 in Figure~\ref{fig:comic} shows the \cd{@Sql} annotation for the the SQL in/out types  derived from the database dictionary.
\end{example}

If the in/out types cannot be determined, a \cd{@SqlUnsupported} annotation is added.
The \cd{@RetrievesSqlResultSet} annotation is used to mark methods that execute a statement and convert it to a result set.
\opsc{} propagates the \cd{@Sql} or \cd{@SqlUnsupported} type information to the result value invocations of methods with this annotation.
The result set only contains results, so only the out types of the \cd{@Sql} annotation are kept.
If the receiver type is \cd{@SqlUnsupported}, then, in sound type checking, \opsc{} issues an error; in degraded type checking, such invocations are quietly ignored.
We annotated the \cd{getResultSet} method of the \cd{Statement} interface and the \cd{executeQuery} method of the \cd{PreparedStatement} interface in the JDBC library with \cd{@RetrievesSqlResultSet}.

\subsubsection{Flow-sensitive type refinement}\label{sec:type-refinement}
The type introduction rules described %
annotate the \emph{results} of method calls that create a statement or result set.
There are also certain method calls where the \emph{receiver} of the call has to be annotated.
Consider the following example, which uses method \cd{stmt.execute()} (returns a \cd{boolean}) instead of the aforementioned \cd{stmt.executeQuery()} (returns a \cd{ResultSet}).
\begin{lstlisting}[numbers=left]
Statement stmt = conn.createStatement();
stmt.execute("SELECT total FROM Invoice");|\label{ln:stmt_execute}|
ResultSet rs = stmt.getResultSet();
\end{lstlisting}
To have the SQL type information available when creating the result set, we need to construct a \cd{@Sql} annotation and attach it to \cd{stmt} after the method call in line~\ref{ln:stmt_execute}.
We ensure that annotations are updated after relevant method calls by defining a custom \emph{transfer function} that can track these calls and infer annotations accordingly.
Having defined the \cd{@Sql} type hierarchy and the least upper bound for two types, annotations are propagated through the control flow graph and merged at control flow join points (e.g., after \texttt{if} statements or loops), ensuring accurate type checking even across complex execution paths.

\subsubsection{Verifying setter and getter calls}\label{sec:verifying-calls}

Whenever a getter/setter is called on a \cd{@Sql}-annotated object, the type of the method can then be compared to the expected types stored in the annotation.
The JDBC-specified type mapping configuration is used to decide if the types match (i.e., are JDBC-recommended).
For getter and setter calls on objects annotated with \cd{@SqlUnsupported}, in sound type checking, \opsc{} issues an error; in degraded type checking, such invocations are quietly ignored, as they are either out of scope for \opsc{}
(for CallableStatements) or a warning has already been emitted for unparsable or unextractable statements.

\begin{example}
Continuing with our running example from Figure~\ref{fig:comic}, the third step shows that the setter-call type-checks successfully, unlike the getter-call (last line of code).
\end{example}

We call getter or setter invocations of unannotated methods \emph{non-local accesses}, because this means that they cannot be traced back to the declaration of the statement or result set.

\subsubsection{Declaring Annotations Manually}\label{sec:manual-annos}

In some cases, when a statement is accessed in multiple methods, the analysis of non-local accesses may  nevertheless be enabled by manually annotating a method signature.
A \cd{@Sql} annotation can be added to a parameter or the return type of a method.
In case of an annotated parameter, \opsc{} will
\begin{enumerate*}[label=(\alph*)]
    \item check setter/getter accesses to the parameter within the method according to the written annotation and
    \item require that the argument in calls to the method match (i.e., be a subtype of) the annotated type.
\end{enumerate*}
For methods with an annotated result type, the checker will
\begin{enumerate*}[label=(\roman*)]
    \item verify setter/getter accesses to the object returned by the annotated method and
    \item make sure that the method returns a value of a subtype of the annotated parameter type.
\end{enumerate*}

We illustrate manual annotations in the upcoming Example~\ref{ex:non-local}.

\subsection{Code Gallery}

The examples below illustrate the capabilities of \opsc. Several are based on third-party code that we also analyze in our experiments.

\begin{example}[Control flow] \label{ex:control-flow}
The following code snippet is based on an open source database benchmark~\cite{escadatpcc}. 
The raw SQL statement string can be extracted by the Constant Value Checker, despite it being concatenated (lines~\ref{ln:concat1} and \ref{ln:concat2}).
Some other tools cannot handle such dynamically created SQL statements.
By accessing the data dictionary, \opsc\ can statically resolve the wildcard (line~\ref{ln:wildcard}) in the SQL statement.
Despite the  while-loop and the conditional, 
\opsc\ can correctly check the setter- and getter-calls.

\begin{lstlisting}[numbers=left]
PreparedStatement statement = null;
ResultSet rs = null;

int _d_id = Integer.parseInt((String) obj.get("did"));

while (_li_no < _o_ol_cnt) {
  statement = con.prepareStatement(
    "select * from stock " + |\label{ln:wildcard}\label{ln:concat1}|
    "where s_i_id = ? and s_w_id = ?");|\label{ln:concat2}|
  statement.setInt(1, _li_id);
  statement.setInt(2, _li_s_w_id);
	
  rs = statement.executeQuery();
  rs.next();
	  
  if (_d_id == 1) {
    _s_dist = rs.getString("s_dist_01"); // CHAR(24) |\label{ln:ifelse_columnname}|
  } else if (_d_id == 2) {
    _s_dist = rs.getString("s_dist_02"); // CHAR(24)
  } 
}
\end{lstlisting}

\end{example}

\begin{example}[Reassigning prepared statements]
\label{ex:reassign}
The following code snippet is adapted from the open-source project OSCAR~\cite{oscarpro}, the de-facto reference application in academic research on database applications (e.g., OSCAR is studied in~\cite{DBLP:conf/csmr/MeuriceC14, DBLP:journals/scp/CleveGMMW15,DBLP:conf/qrs/MeuriceNC16,DBLP:conf/csmr/GoeminneDM14,DBLP:conf/edbt/BeineHWC14,8471041,7081881,DBLP:conf/issta/VasquezLVP16}).

\begin{lstlisting}[numbers=left]
String sql = "SELECT image_id FROM client_image "
  + "WHERE image_data IS NOT NULL AND contents IS NULL";
PreparedStatement pst = conn.prepareStatement(sql);

sql = "SELECT image_data FROM client_image  "
  + "WHERE image_id = ?";
pst = conn.prepareStatement(sql);

pst.setLong(1, id); 
\end{lstlisting}

Variable \cd{pst} is first associated with a query that has no parameters and later reassigned to a different query with one parameter. Other tools
commonly cannot track this reassignment along the control flow. For example, when IntelliJ IDEA Ultimate encounters \cd{pst.setLong(1, id)}, it assumes that the statement has no placeholders and issues an error. In contrast, \opsc\ checks the setter against the correct SQL statement. 

Like in the previous example, the prepared statement is reassigned. Yet previously, the reassignment happens inside a while-loop (line~7), and the SQL statement string itself remains the same. Now, the SQL statement string changes between assignments. \opsc\ can reliably handle both scenarios.
\end{example}

\begin{example}[Sequential parameter binding] 
\label{ex:increment}
\opsc{} supports common programming idioms, such as sequential parameter binding. For the code snippet below,
the Constant Value Checker (see Section~\ref{sec:value-checker}) statically determines the index of the setter-calls, even though not supplied as an integer literal. Thus, \opsc\ can check the setter-calls with index \cd{ctr++}, 
while other developer tools 
are not able to track the incremented index through the control flow.

\begin{lstlisting}[numbers=left]
PreparedStatement ps = conn.prepareStatement("...");
int ctr = 1;
ps1.setInt(ctr++, quantity); %
ps1.setString(ctr++, id);    %
\end{lstlisting}
\end{example}

\begin{example}[Non-local type checking]\label{ex:non-local}
The following code is based on a repository for Java design patterns~\cite{java-design-patterns}.

\begin{lstlisting}[numbers=left]
public Optional<Room> getById(int id) throws Exception {
  //...
  var statement = connection.prepareStatement(
    "SELECT * FROM ROOMS WHERE ID = ?");
  statement.setInt(1, id);
  resultSet = statement.executeQuery();
  if (resultSet.next()) {
    return Optional.of(createRoom(resultSet));
  }
  // ...
}

private Room createRoom(
  ResultSet resultSet
) throws Exception {
  return new Room(
    #// calls getInt instead of getBigDecimal#  
    #resultSet.getInt("ID")#,
    resultSet.getString("ROOM_TYPE"),
    resultSet.getInt("PRICE"),
    #// calls getBoolean instead of getString#
    #resultSet.getBoolean("BOOKED")#);
}
\end{lstlisting}

Method \verb!getByID(id)! fetches the details of a given hotel room based on the room number. It then calls helper method \verb!createRoom! that instantiates an instance of class \verb!Room!, based on the query result.

Out of the box, \opsc\ cannot type-check these non-local accesses (see previous section).
Yet in lines~14 and~15, 
the signature of the helper method has been manually annotated with the schema of the \verb!ResultSet! that is passed as a parameter.
This enables \opsc\ to statically check the getter-calls (lines~20--24): While all column names exist, not all types match.

\opsc\ will statically check that the query result (or rather, the \verb!ResultSet!) for the query in line~4 matches the annotation for the helper method \verb!createRoom!, when called in line~8.
Calling the method with a \cd{ResultSet} that does not contain the columns specified in the annotation would cause a compilation time error.

\end{example}

\begin{example}[Limitations] \label{ex:limitation}
In the generic helper method below, the setter-call is a non-local access.
Here, the index of the parameter required to be compatible with \cd{setString} depends on the method parameter \cd{parameterIndex}. Thus, a manual \cd{@Sql} annotation cannot resolve this.\footnote{More expressive dependent types could be used to express this relationship, but we leave this additional complexity for future work, if we see a clearer need for it.} 
\opsc\ cannot check this statement, and reports an error in sound mode, or ignores the setter-call in degraded mode.

\begin{lstlisting}[numbers=left]
public void bindParam(PreparedStatement ps,
  int parameterIndex, String value) throws SQLException {
    ps.setString(parameterIndex, value);
}
\end{lstlisting}
\end{example}

\emph{Improvements over state-of-the-art.}
The code snippet below contains three bugs in setting parameters in a prepared statement. Each causes a runtime error.
Existing developer tools like IntelliJ only detect the third error (line~\ref{lst:insertGenre3}) (by counting the number of question marks), while \opsc\ detects all errors.

\begin{lstlisting}[numbers=left]
public void insertGenre(boolean newInstance) 
  throws SQLException {
  String stmt;
  stmt = "INSERT INTO genre (id, name) VALUES (?,?)";

  PreparedStatement ps = conn.prepareStatement(stmt);

  #// Assignments use wrong index#
  #ps.setString(1, "scary industrial hip hop");# |\label{lst:insertGenre1}|
  #ps.setInt(2, 1);# |\label{lst:insertGenre2}|
  #ps.setString(3, "hip hop"); // Index out of bounds# |\label{lst:insertGenre3}|
}
\end{lstlisting}

In this code gallery, we studied small code snippets that illustrate the capabilities of \opsc.  In our upcoming experiments, we will put \opsc\ to the test and  analyze full-sized code repositories.

\section{Experiments}
\label{sec:experiments}

\subsection{Research Hypotheses}

Our experiments explore the hypotheses listed below:
Hypothesis~H1 explores \opsc\ in sound mode, Hypothesis~H2 explores the degraded mode.
A high number of false positives would render \opsc\ unusable in practice (H2.1). Moreover, if the manual annotation effort is excessive or is not effective in increasing the reach of the analysis, the effort is not justified~(H2.2). Finally, to be integrated into the developer tool chain, code compilation with \opsc\ must not impose unnecessary overhead~(H2.3).

\begin{itemize}
\item[\textbf{H1}] \opsc{} is sound; therefore, it produces no false negatives.

\item[\textbf{H2}]
Type-checking existing JDBC code repositories with \opsc{} in degraded mode is practical:
\begin{itemize}
    \item[\textbf{H2.1}] \opsc{} does not excessively produce false positives. 
    \item[\textbf{H2.2}] Manual code annotation increases the reach of static analysis, compared with an unannotated baseline. 
    \item[\textbf{H2.3}] The overhead during code compilation is reasonable. 
    
\end{itemize}
\end{itemize}

Existing developer tools can reliably detect malformed SQL statement strings in JDBC application code.
Therefore, our experiments exclusively target the getter/setter-related exceptions (specifically, the scenarios~(2)-(4) in Proposition~\ref{prop:1}).

\subsection{Setup}
\label{sec:experiments_setup}

\emph{Implementation.}
We implemented \opsc{} based on the EISOP Checker Framework~\cite{eisop-checkerframework} (version~\emph{3.49.5-eisop1}).
Our checker extension is lightweight and comprises only approx.~3.0k
lines of Java code.

\opsc\ can check Java~17+ and Java~8 applications. Specifically, we use OpenJDK 8u482 to compile (and check) the code repository of the OSCAR project (to be introduced below), and OpenJDK 17.0.18+8 for all other code repositories in our experiments.

We implemented type-checking for all JDBC getter- and setter-calls,
with the exception of \cd{setObject}, since we did not encounter such calls in the analyzed third-party code.
Furthermore, our \opsc\ prototype does not cover type-checking for calls to stored procedures or for automatically generated keys. 
All of these limitations could be easily addressed in productization.

\opsc\ supports PostgreSQL with JDBC driver version~42.7.5,
and provides limited support for MySQL and MySQL Connector/J~8.2.0.

In our experiments, we consider a SQL statement string \emph{well-formed} if it can be statically extracted from the code (using Constant Value Checker) and successfully validated against the data dictionary using Calcite.
To do so, \opsc{} requires a library for parsing and type-checking SQL queries. 
We delegate this task primarily to Apache Calcite~\cite{calcite} (version 1.37.0).
Calcite supports a wide variety of SQL dialects, but not all PostgreSQL query constructs.
As fallback for unsupported constructs, we request metadata via JDBC.%
\footnote{
The PostgreSQL JDBC driver again provides more type information compared to MySQL's. However, it does not provide the names of result columns and only limited information about \cd{PreparedStatement} parameter types.}

\emph{Competitor Software.}
We compare against the professional IDE %
IntelliJ IDEA Ultimate 2025.1.5.1, with Plugin ``Database Tools and SQL'' (bundled 251.28293.39).

\emph{Execution Environments.}
Unless stated otherwise, all experiments were conducted on a developer notebook (MacBook Air M3 with 16~GB of memory).  In measuring overheads during Java compilation, we use a Linux server equipped with two 3.1~GHz Intel Xeon Gold processors and 384~GB of main memory as build server.

\begin{table*}[]

    \caption{Analyzed code repositories, including handwritten and third-party code (T: from textbooks, RW: real-world applications, BM: database benchmark). Stating lines of code (LoC), number of tables in the schema (\#Tables), and number of JDBC {\tt Statement}s and {\tt PreparedStatement}s  (with max.\ number of parameters). Listing max.\ SQL string length and number of getters/setters. Finally, share of SQL statement strings that are well-formed w.r.t.\ the database schema.}
    \label{tab:data}

    \centering
    \small

\setlength{\aboverulesep}{0.2ex}
\setlength{\belowrulesep}{0.2ex}
\setlength{\abovetopsep}{0.2ex}
\setlength{\belowbottomsep}{0.2ex}
\renewcommand{\arraystretch}{0.9} %

\begin{tabular}{l@{\hspace{4pt}}rlr@{\hspace{4pt}}rrrr@{\hspace{4pt}}rrrr}
    \toprule
    Repository &  & Commit & LoC (k) & \#Tables & DynStmt & PrepStmt & MaxParams & MaxSQLLen & getters & setters & WellFormedSQL \\
    \midrule      

    Handwritten test suite
    & &
    & \mbox{1\phantom{*}} & 11 & 2 & 43 & 4 & 99 & 18 & 53 & 45 (100\%)
    \\
    \midrule

    T/O'Reilly: bank & \cite{reese2000database}  & 
    \href{https://resources.oreilly.com/examples/9781565926165/-/tree/6ca73dfc6fbd43e4fe37294255e6614815a18b6c/examples/etc}{\cd{6ca73df}} & 
    \mbox{6\phantom{*}}
    & 4 & 0 & 14 & 9 & 146 & 16 & 31 & 11 (\phantom{1}79\%)
    \\    

    T/java-design-patterns & \cite{java-design-patterns} &
    \href{https://github.com/iluwatar/java-design-patterns/tree/163c3017bb356937d876cd9a05905c012f3b0af6}{\cd{163c301}}
    & 
    5$^*$ & 8 & 14 & 25 & 5 & 80 & 14 & 41 & 26 (100\%) \\
    
    T/JDBC-Course & \cite{jdbc-course} & 
    \href{https://github.com/sayedabdul-aziz/JDBC-Course/tree/04ed1613c612f8d9ae53ef7629c3cb254d6cad40}{\cd{04ed161}}
    & 
    \mbox{1\phantom{*}} & 3 & 12 & 5 & 3 & 57 & 9 & 14 & 13 (\phantom{1}81\%) \\

    \midrule
    
    RW/OSCAR & \cite{oscarpro} &
    \href{https://bitbucket.org/oscaremr/oscar/src/cca70ec9a265370992a8f55d5bcb82d011c4b6ac/}{\cd{cca70ec}} &
    \mbox{852\phantom{*}} & 558 & 54 & 97 & 29 & 1402 & 166 & 197 & 56 (\phantom{1}37\%) \\

    RW/OpenNMS & \cite{opennms} & 
    \href{https://github.com/OpenNMS/opennms/tree/dcafd6dcd0c4f5e2d5219091cb74f5190d56f309}{\cd{dcafd6d}} & 
    64$^*$ & 120 & 10 & 52 & 63 & 1875 & 29 & 178 & 34 (\phantom{1}55\%) \\

    \midrule

    BM/EscadaTPC-C & \cite{escadatpcc} &
    \href{https://github.com/rmpvilaca/EscadaTPC-C/tree/ec47ca014fc31bb81f5183593c757c34508d3820}{\cd{ec47ca0}} &
    2$^*$ & 9 & 4 & 51 & 21 & 268 & 72 & 190 & 46 (\phantom{1}88\%) \\

    \bottomrule
\end{tabular}
\end{table*}

\subsection{Analyzed Code}
\label{sec:experiments_code}

Table~\ref{tab:data} lists the code repositories analyzed.
The first is a handwritten test suite which includes small programs, in the style of Examples~\ref{ex:problem_getters}, \ref{ex:problem_setters},  and~\ref{ex:increment}. All SQL statement strings are well-formed.
The tests comprise 31 positives and 40 negatives.

The remaining repositories contain functional third-party Java code and a database schema. %
One group~(T) consists of code from textbooks and tutorials. This code is small, with simple control flow, and showcases a wide range of JDBC features.
Group~(RW) aims at testing the practicality of using \opsc{} in larger-sized, real-world application code. This group includes
 OSCAR,
 an electronic medical record system, and OpenNMS, a network monitoring platform.
Both applications have non-trivial control flow and have been actively maintained for over~25 years.
Last is a Java implementation of the TPC-C database benchmark~(BM).  
We analyze its implementation for MySQL, which includes SQL queries that use wildcards, as illustrated in Example~\ref{ex:control-flow}.

\emph{Summary Statistics.}
In Table~\ref{tab:data}, we state the repository name, origin, and version (git commit hash). 
We also state its size in lines of code. We exclude directories with unrelated code (not using JDBC), and indicate this by an asterisk.
We state the size of the database schema in terms of the number of tables. %

Regarding SQL, we consider CRUD-statements only. We include WITH in queries, but exclude WITH RECURSIVE (which does not appear in any of the third-party projects), as well as DDL statements, such as DROP TABLE.
Again, these limitations are only technical.

We state the total number of declarations of JDBC \texttt{Statement} and \texttt{PreparedStatement}, as well as the maximum number of parameters in prepared statements.
In OpenNMS, this exceeds 60 parameters in a single statement.
We state the maximal lengths of SQL statement strings. The very long statements in OSCAR and OpenNMS are commonly INSERTs with many parameters.
We state the total numbers of getter- and setter calls. 

We state the share of well-formed SQL statement strings.
Interestingly, we found malformed SQL strings in the textbook code.\footnote{Listing~\ref{lst:oreilly-nonparsable} illustrates an example from the textbook collection, where the number of column names and question marks are off by one.
Meanwhile, we have confirmed this problem with the author.
In \emph{java-design-patterns}, six statements are malformed.
For four, we could not find any matching SQL dialect, which suggests a bug.
Two other statements contain the MySQL-specific \cd{SHOW COLUMNS} clause.}
In the real-world projects, not all SQL commands are recognized by Calcite. 
Since OSCAR uses MySQL-proprietary SQL syntax, yet \opsc{} only provides limited support for MySQL, the share of well-formed SQL statements is lowest for OSCAR. 

\emph{Ground truth.}
With our handwritten test suite, we can test \opsc\ against a known ground truth.
However, in analyzing functional third-party code, we cannot expect to find bugs that cause runtime exceptions. However, we were able to identify an older version of the database benchmark repository with a bug in a getter-call that was fixed in a later commit.\footnote{In the commit \href{https://github.com/rmpvilaca/EscadaTPC-C/tree/b3f8f8d4ad36f238c02fda7e1b5726494ede935b}{\cd{b3f8f8d}}, a \cd{setString} method call is changed to \cd{setInt}.} 
Therefore we choose this specific, older version of \emph{EscadaTPC-C} for our analysis.

\emph{SQL dialects.}
We made minor code changes to account for SQL dialects:
For OSCAR, 
this amounts to 5 lines of Java code out of over 800K, and 45 out of 19K lines of the MySQL schema definition.%
\footnote{
To provide an idea how small-scale these changes are, we describe two:
We replaced MySQL-proprietary data type \cd{TEXT} with \cd{VARCHAR} in the data dictionary.
Because it is a reserved keyword in Calcite, we escaped a column named \cd{datetime} in two locations.}
We also changed~5 lines in the \emph{java-design-patterns} repository.\footnote{
This code uses a \cd{BLOB} type
which is not supported by PostgreSQL.
As a workaround, we changed the type of a single database column from \cd{BLOB} to PostgreSQL \cd{BYTEA} and adjusted one setter- and one getter-call accordingly.}

\subsection{Soundness}

In exploring hypothesis~H1, we analyze the handwritten test suite. 

\paragraph{Setup}
We analyze the code as is and compare it with IntelliJ. %

\begin{figure}[tb]
\centering
\subcaptionbox{\opsc\label{fig:cm-ours}}[0.45\linewidth]{
\footnotesize
\centering
\begin{tabular}{@{}lcc@{}}
\toprule
 & Pred.~Pos & Pred.~Neg \\
\midrule
Actual Pos & TP: 31 & FN:  0 \\   
Actual Neg & FP:  0 & TN: 40 \\
\bottomrule
\end{tabular}
}
\hfill
\subcaptionbox{IntelliJ\label{fig:cm-intellij}}[0.45\linewidth]{
\footnotesize
\centering
\begin{tabular}{@{}lcc@{}}
\toprule
 & Pred.~Pos & Pred.~Neg \\
\midrule
Act. Pos & TP: 6 & FN: 25 \\   
Act. Neg & FP: 0 & TN: 40 \\
\bottomrule
\end{tabular}
}
\caption{Confusion matrices for handwritten test suite: Predicted vs. actual positives/negatives, (a)~\opsc{} vs.\ (b)~IntelliJ.}

\label{fig:confusion-matrices}
\end{figure}

\emph{Results.}
Figure~\ref{fig:cm-ours} shows the confusion matrix for \opsc.
\opsc\ produces no false negatives. %
IntelliJ (Figure~\ref{fig:cm-intellij}) correctly detects 6 cases when a setter accesses a parameter index position that is ``out of bounds'', yet it leaves 25 false negatives: It does not detect when a getter attempts to access a non-existent column (whether by name or index), nor any getter/setter type mismatches. 

\emph{Discussion.}
\textbf{\opsc{} is sound and therefore does not produce false negatives}. Our experiments are in line with hypothesis~H1 and also show that \opsc{} improves over the state-of-the-art.

\subsection{Ad-hoc Code Analysis}
\label{sec:ex_adhoc}

We now explore the out-of-the-box experience with \opsc{} applied to the third-party code repositories.

\emph{Setup.} We analyze the third-party code repositories \emph{ad hoc}, without  manual annotations. 
\opsc{} runs in degraded mode. We report the following cases for local getter/setter accesses:
\begin{enumerate*}
    \item Positive: \opsc{} reports a type mismatch.
    \item Negative: \opsc{} reports no type mismatch.
\end{enumerate*}
We manually distinguish true and false positives.
We refer to  non-local accesses as \emph{out-of-scope} (abbreviated OOS).

We further analyze the same code repositories with IntelliJ. In analyzing OSCAR, IntelliJ ran into timeouts trying to analyze the entire repository, so we checked file-by-file.

\emph{Results.}
Table~\ref{tab:adhoc_textbook} summarizes the results for the textbook code. 
\opsc{} produces no false positives~(FP). 
For two repositories, 
\opsc{} actually detects true positives~(TP). Upon inspection, the true positives seem low-risk.
For example, in \emph{JDBC-Course}, \cd{getString} is called on a SQL \cd{INTEGER}.
Some accesses are not local (denoted OOS) and cannot be checked ad hoc.

\begin{table}[tb]
    \caption{Ad-hoc analysis of textbook code (degraded mode, stating true/false positive/negatives, and out-of-scope).}
    \label{tab:adhoc_textbook}

    \centering
    \small

\setlength{\aboverulesep}{0.2ex}
\setlength{\belowrulesep}{0.2ex}
\setlength{\abovetopsep}{0.2ex}
\setlength{\belowbottomsep}{0.2ex}
\renewcommand{\arraystretch}{0.9} %
    
    \begin{tabular}{l rrr r}
    \toprule
    Project & TP & FP & N & OOS
    \\
    \midrule 
     O'Reilly: bank &  0  & 0 & 47 & 1  \\
     java-design-patterns & 11 &0 & 44 & 21  \\
     JDBC-Course & 2 & 0 & 21 & 0 
         \\

    \bottomrule 
    \end{tabular}

\end{table}

\begin{figure}[tb]
  \centering
  \begin{subfigure}[t]{0.55\columnwidth}
    \centering
     
    \def\xText{TP,FP,N,OOS}
\newcounter{TextA}

\psset{xunit=.44cm,yunit=.012cm,ticksize=-3pt 3pt}

\begin{pspicture}(-2,-35)(8,350)

\footnotesize

\psaxes[axesstyle=axes,Dy=50,Dx=2,labels=y](4,280)

\rput(0,280.4){\textbf{\huge$\approx$}}
\rput(0,285){%
  \psaxes[xAxis=false,Dy=50,Oy=500,Dx=2](4,50)
}
\psset{interrupt={280,5,220}}

\listplot[linecolor=black,plotstyle=bar,barwidth=0.3cm,
  fillcolor=black!35,fillstyle=solid]{
    2   0   %
    3 215   %
    4 527   %
}

\uput{4pt}[90](2,1){0}

\rput{90}(-2.3,170){Frequency}

\newcounter{stackheightA}
\newcommand{\stackbar}[3]{%
  \listplot[linecolor=black,plotstyle=bar,barwidth=0.3cm,
    fillcolor=#3,fillstyle=solid]{ #1 \thestackheightA }%
  \addtocounter{stackheightA}{-#2}%
}

\setcounter{stackheightA}{148}
\stackbar{1}{36}{cb1}
\stackbar{1}{30}{cb2}
\stackbar{1}{27}{cb3}
\stackbar{1}{22}{cb4}
\stackbar{1}{12}{cb5}
\stackbar{1}{21}{cb6}

\psforeach{\nA}{\xText}{%
  \stepcounter{TextA}
  \rput[rc]{90}(\theTextA,-4mm){\nA}
}

\newcommand{\legendentry}[2]{\raisebox{.7\height}{\psframebox[fillstyle=solid,fillcolor=#1]{}} & {\footnotesize #2} }%

\rput(4.7,120){%
  \psframebox[fillstyle=solid,fillcolor=white,framesep=2pt]{%
    \begin{tabular}{@{} l @{\hskip 3pt} l @{}}
      \legendentry{cb1}{getString/INTEGER} \\
      \legendentry{cb2}{setLong/INTEGER} \\
      \legendentry{cb3}{getString/BIGINT} \\
      \legendentry{cb4}{setString/INTEGER} \\
      \legendentry{cb5}{getLong/INTEGER} \\
      \legendentry{cb6}{Other}
    \end{tabular}%
  }%
}

\end{pspicture}
    \caption{OSCAR}
    \label{fig:class-oscar}
  \end{subfigure}
  \begin{subfigure}[t]{0.43\columnwidth}
    \centering
    \vspace{5pt}

    \begin{subfigure}[t]{\columnwidth}
      \centering
      \def\xText{TP,FP,N,OOS}
\newcounter{TextB}

\psset{xunit=.44cm,yunit=.009cm,ticksize=-3pt 3pt}

\begin{pspicture}(-2,-10)(4.8,148)

\footnotesize

\psaxes[axesstyle=axes,Dy=25,Dx=2,labels=y](4,49)

\rput(0,49.4){\textbf{\huge$\approx$}}
\rput(0,49){%
  \psaxes[xAxis=false,Dy=50,Oy=125,Dx=2](4,100)
}
\psset{interrupt={49,5,75}}

\listplot[linecolor=black,plotstyle=bar,barwidth=0.3cm,
  fillcolor=black!35,fillstyle=solid]{
    2   0   %
    3 203   %
    4 151   %
}

\uput{4pt}[90](2,0){0}
\uput{4pt}[90](1,4){4}

\newcounter{stackheightB}
\newcommand{\stackbar}[2]{%
  \listplot[linecolor=black,plotstyle=bar,barwidth=0.3cm,
    fillcolor=black!35,fillstyle=solid]{ #1 \thestackheightB }%
  \addtocounter{stackheightB}{-#2}%
}

\setcounter{stackheightB}{4}
\stackbar{1}{4}

\psforeach{\nA}{\xText}{%
  \stepcounter{TextB}
  \rput[rc]{90}(\theTextB,-4mm){\nA}
}

\end{pspicture}
        \vspace{10pt}
      \caption{OpenNMS}
      \label{fig:b}
    \end{subfigure}

    \vspace{13pt}

    \begin{subfigure}[t]{\columnwidth}
      \centering
      \def\xText{TP,FP,N,OOS}
\newcounter{TextD}

\psset{xunit=.44cm,yunit=.008cm,ticksize=-3pt 3pt}

\begin{pspicture}(-2,-10)(4.8,148)

\footnotesize

\psaxes[axesstyle=axes,Dy=50,Dx=2,labels=y](4,99)

\rput(0,99.4){\textbf{\huge$\approx$}}
\rput(0,105){%
  \psaxes[xAxis=false,Dy=50,Oy=200,Dx=2](4,50)
}
\psset{interrupt={100,5,100}}

\listplot[linecolor=black,plotstyle=bar,barwidth=0.3cm,
  fillcolor=black!35,fillstyle=solid]{
    1  30   %
    2   0   %
    3 232   %
    4   0   %
}

\uput{4pt}[90](2,0){0}
\uput{4pt}[90](4,0){0}

\psforeach{\nA}{\xText}{%
  \stepcounter{TextD}
  \rput[rc]{90}(\theTextD,-4mm){\nA}
}

\end{pspicture}
       \vspace{10pt}
      \caption{EscadaTPC-C}
      \label{fig:d}
    \end{subfigure}

  \end{subfigure}

  \caption{Ad-hoc analysis of functional software: 
  Manually confirmed true positives (TP) and (zero) false positives (FP) demonstrate practical usability. 
  Also reporting accesses classified as negatives (N) and  non-local accesses (OOS). 
  }
  \label{fig:ad-hoc-analysis}

\end{figure}
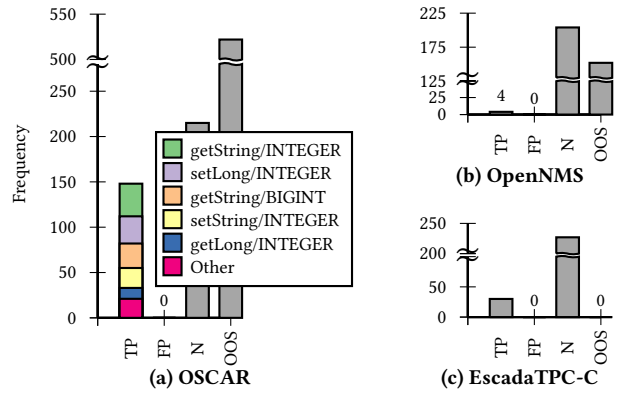

Figure~\ref{fig:ad-hoc-analysis} shows the results for the remaining repositories. The visualization via bar charts provides a sense of scale, and we highlight the most frequent true positives for OSCAR inside the legend.
Again, \opsc{} produces no false positives. 

For OSCAR, \opsc{} reports approx.~150 positives, which is still in the range that a developer can check one-by-one.
The most frequent true positive is calling \cd{getString} on a ResultSet column of SQL-type \cd{INTEGER} (like with the textbook code above), which is not likely to be critical in production.
As may be expected, the share of get\-ters/set\-ters classified as ``negatives'' is higher.

OSCAR has a considerable share of non-local accesses that could not be checked. 
In OpenNMS, the share of non-local accesses is smaller. In the benchmark application, there are none at all.

For our analysis with IntelliJ, we only report on OSCAR and OpenNMS; for the other projects, the commercial tool does not detect problems beyond malformed SQL statements, which is not the focus here.
Every true positive recognized by IntelliJ regarding getters/setters is also recognized by \opsc.
IntelliJ produces one false positive, by not properly tracking the control flow (this scenario is shown in Example~\ref{ex:reassign}).
Again, the analysis with IntelliJ does not detect any type mismatches in getters/setters.

\emph{Discussion.}
The presence of true positives in functional code surprises, yet inspection reveals that they mostly concern lower-risk type mismatches. In degraded mode, it therefore makes sense to enable developers to configure different warning and error levels.  

However, \opsc{} also found access calls that can indeed cause runtime exceptions:
\begin{enumerate*}
\item     
One true positive is the confirmed bug in \emph{EscadaTPC-C}, which originally motivated us to choose this specific code version for analysis (discussed in Section~\ref{sec:experiments_code}).
 Thus, \textbf{\opsc{} found a confirmed bug}.

\item 
In  \emph{OpenNMS}, \opsc{} found getter-calls where the column name does not exist. When executed, this would cause runtime exceptions. 
As these calls are within deprecated classes, inside a conditional branch that is currently not executed, these problems could go undetected so far. 
Thus, \textbf{\opsc\ found undetected bugs lurking in dead code}.
\end{enumerate*}
Making developers aware of such risks in their codebase is a valuable contribution.

 \textbf{Here, \opsc{} did not produce any false positives}, confirming Hypothesis~H2.1. Consequently,  developers are unlikely to be overwhelmed by unfounded messages.

The high share of reported negatives is to be expected as we analyze functional code. Overall, our results strongly indicate that \textbf{developers are likely to find \opsc{} useful out-of-the-box}:
Compared to tools like IntelliJ, \opsc\ finds important errors that existing tools cannot detect.

What stands out is the substantial share of non-local accesses in OSCAR, unlike with OpenNMS and the benchmark repository. 
We suspect that this is due to project-specific programming patterns.

\subsection{Manual Annotations}

Hypothesis H2.2 concerns the effort of manually annotating code.

\emph{Setup.}
As mentioned earlier, manual annotations may be required for the intra-procedural case where statements or result sets are passed across method call boundaries.
We added manual annotations to the code of OSCAR and OpenNMS as follows.  
For each case where SQL statement handling spanned over several methods, we inspected the code to determine if the invocations can be described by fixed SQL types.
Where this was the case, we added annotations. 
As a proxy metric for the annotation effort, we measure the string lengths of annotations (excluding whitespaces).

We analyze the code thus annotated with \opsc{} running in degraded mode.
We count the getters/setters that can be type-checked ad-hoc, as well as those that can be type-checked due to manual annotation, and those that nevertheless remain out of scope.

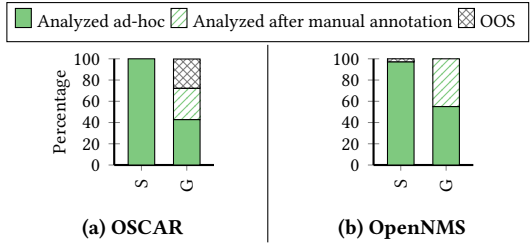
\begin{figure}[tb]
  \centering
  \ref{coverage-legend} %

  \begin{subfigure}[t]{0.40\columnwidth}
    \centering
    \begin{tikzpicture}[remember picture]

\begin{axis}[
    x=0.6cm,
    y=0.4pt,
    nodes={font=\footnotesize},
    axis line style={line width=0.8pt},
    ylabel={Percentage},
    y label style={at={(-0.36,0.5)}},%
    clip=false,
    height=5.5cm,
    xmin={[normalized]-0.6},
    xmax={[normalized]1.6},
    ymin=0,
    y tick style={line width=0.3pt},
    ybar stacked,
    bar width=10pt,
    symbolic x coords={OSCAR setters, OSCAR getters},
    xticklabels={S, G},
    xtick=data,
    x tick label style={rotate=90},
    axis lines=left,
    cycle list={%
        {fill=cb1},
        {pattern color=cb1,pattern={north east lines}},
        {pattern color=Gray,pattern={crosshatch}},
        {fill=cb4},
        {fill=cb5},
        {fill=cb6}
    },
    legend style={
      font=\footnotesize, 
    },
    legend columns=-1,
    legend entries={Analyzed ad-hoc,Analyzed after manual annotation,OOS},
    legend to name={coverage-legend},
]

\addplot+[ybar] plot coordinates {(OSCAR setters,100) (OSCAR getters,42.7)}; %
\addplot+[ybar] plot coordinates {(OSCAR setters,000) (OSCAR getters,29.5)}; %
\addplot+[ybar] plot coordinates {(OSCAR setters,000) (OSCAR getters,27.6)}; %

\end{axis}

\end{tikzpicture}
    \caption{OSCAR}
    \label{fig:coverage-oscar}
  \end{subfigure}
  \rulesep
  \begin{subfigure}[t]{0.4\columnwidth}
    \centering
    \begin{tikzpicture}[remember picture]
\begin{axis}[
    x=0.6cm,
    y=0.4pt,
    nodes={font=\footnotesize},
    axis line style={line width=0.8pt},
    clip=false,
    height=5.5cm,
    xmin={[normalized]-0.6},
    xmax={[normalized]1.6},
    ymin=0,
    ybar stacked,
    y tick style={line width=0.3pt},
    bar width=10pt,
    symbolic x coords={OpenNMS setters, OpenNMS getters},
    xticklabels={S, G},
    xtick=data,
    x tick label style={rotate=90},
    axis lines=left,
    cycle list={%
        {fill=cb1},
        {pattern color=cb1,pattern={north east lines}},
        {pattern color=Gray,pattern={crosshatch}},
        {fill=cb4},
        {fill=cb5},
        {fill=cb6}
    },
    legend style={
      font=\footnotesize, 
      /tikz/every node/.style={anchor=west, align=left},
      at={(0.97,0.4)}
    },
    reverse legend
]

\addplot+[ybar] plot coordinates {(OpenNMS setters,97) (OpenNMS getters,55)}; %
\addplot+[ybar] plot coordinates {(OpenNMS setters,00) (OpenNMS getters,45)}; %
\addplot+[ybar] plot coordinates {(OpenNMS setters,03) (OpenNMS getters,00)}; %

\end{axis}

\coordinate (legend) at (0.4, 2.75);

\end{tikzpicture}
    \caption{OpenNMS}
    \label{fig:coverage-opennms}
  \end{subfigure}

  \caption{
    \opsc{} in degraded mode. Share of analyzable setters~(S) and getters~(G) that can be type-checked ad-hoc vs.\
     after manual annotation, or remain
    out of scope (OOS).
    }
    
    \label{fig:coverage-bars}
\end{figure}

\emph{Results.}
The number of code locations requiring annotation is small, only~5 for OSCAR and~6 for OpenNMS. The annotation strings are long, reaching approx.\ 1.9k characters for OSCAR and~1.3k characters for OpenNMS.
Yet, they are not complex:
OSCAR uses methods that handle query results from tables with many columns, which must all be listed in the annotations.
Annotations could in many cases be copied directly from type information reported in \opsc's error messages for the program without annotations.

Figure~\ref{fig:coverage-bars} visualizes the share of getters/setters that can be type-checked.
For OSCAR, all setter-calls can be type-checked ad-hoc, while
more than half of the getter-calls are out of scope for an ad-hoc analysis. Half of these cases type-check with manual annotations.

More than half of the OpenNMS setter/getter calls can be type-checked ad-hoc. With annotations, \opsc{} type-checks close to~100\%.

For OSCAR, the additionally type-checked getters/setters do not produce new true positives. For OpenNMS, we identify 17 additional true positives. \opsc{} still reports no false positives. 

\emph{Discussion.}
In quantifying the annotation effort, we count the annotated code locations and lines of code. This is only  a crude proxy metric to the time it will takes developers to write the manual annotations.
We were able to come up with the required annotations by a straightforward inspection of the method call hierarchy and copying from error messages of \opsc's checker. While no guarantee, these observations indicate that there is potential for a (semi-)automatic inter-procedural type inference engine for SQL types that can produce a large part of required annotations.

Clearly, the \textbf{reach of  \opsc{} depends on how programmers modularize their code}: With OSCAR, about 75\% of getters can be checked, yet with OpenNMS, it is nearly all getters/setters.  

All in all, we consider the \textbf{manual annotation effort to be very reasonable}, given the potential to \textbf{significantly increase the number of getters/setters that can be type-checked}. This confirms Hypothesis~H2.2.

\subsection{\opsc{} Overheads}%

Finally, we monitor the overheads that static analysis with \opsc\ adds to the compilation of the Java code.

\emph{Setup.}
We compile the two real-world projects on the build server.
As performance baselines, we compile the code with (1)~the Java compiler, (2)~the Checker Framework with the Value Checker plugin (which also invokes the Java compiler, see Section~\ref{sec:prelims_checker}) and (3)~with \opsc{} that also relies on the Checker Framework and the compiler.

We measure the run time of the compilation using Linux \verb!time!, and the maximum memory consumption during compilation using \verb!/usr/bin/time!. For each metric, we perform~5 consecutive runs, discard the minimum and maximum measured values, and compute the mean of the remaining three.

\begin{table}[t]

\caption{Compilation time and maximum memory usage during code compilation (mean values) for the Java compiler~(JC), the value checker~(VC) and the \opsc{} checker}
\label{tab:overheads}

\centering
\small

\setlength{\aboverulesep}{0.2ex}
\setlength{\belowrulesep}{0.2ex}
\setlength{\abovetopsep}{0.2ex}
\setlength{\belowbottomsep}{0.2ex}
\renewcommand{\arraystretch}{0.9} %

\begin{tabular}{l l rrr}
\toprule
Metric & Repository & JC & JC+VC & JC+VC+\opsc \\
\midrule
\multirow{2}{*}{Comp.\ Time (s)}
    & OSCAR   & 43  & 340 & 515 \\
    & OpenNMS & 15  & 31  & 43  \\
\midrule
\multirow{2}{*}{Memory (MB)}
    & OSCAR   & 2,751 & 3,844 & 4,467 \\
    & OpenNMS & 2,076 & 2,153 & 2,133 \\
\bottomrule
\end{tabular}

\end{table}

\emph{Results.}
We first consider the run time of compilation in Table~\ref{tab:overheads}.
\opsc{} takes under 9 minutes to compile OSCAR's over 850k lines of code. This is about 12~times slower than the plain vanilla Java compiler, yet compared to the Checker Framework baseline, only a slowdown by a factor of~1.5.
For OpenNMS, which has a much smaller code base, the gap between the Java compiler and \opsc{} is only a slowdown by factor of~2.9.

We next consider the  maximum memory usage during code compilation. The memory usage varies between runs, and we report mean values. The difference between checking with \opsc\ versus checking with the ValueChecker only is a factor of ca.~1.2.
With OpenNMS, these effects are less pronounced.

\emph{Discussion.}
Compiling OSCAR with \opsc{} enabled takes too long for interactive software development, but \textbf{compilation time is acceptable to be part of continuous builds}, and in any case, nightly builds. 
This confirms Hypothesis~2.3. 
Since \opsc\ is not coupled to an interactive IDE, but is part of the Java compilation, \textbf{integration into the build chain is easily feasible}.

The memory usage for compiling OSCAR with \opsc\ is noticeably higher than the compilation with the Java compiler. Yet for a well-equipped build server, this does not constitute a problem. Moreover, all preceding experiments were successfully conducted using only a developer notebook.

Overall, OSCAR as the largest project analyzed stands out w.r.t.\ compilation overhead.
Our manual investigation of the code reveals that OSCAR contains classes with many static strings, which is also demanding on the Value Checker. 
There is further evidence indicating that OSCAR is an inherently challenging project to analyze: As pointed out in Section~\ref{sec:ex_adhoc}, analyzing the entire OSCAR project with IntelliJ on the developer laptop failed due to timeouts, and we had to resort to checking the project one file at-a-time.

\section{Discussion}
\label{sec:discussion}

Our experiments with \opsc\ in sound mode demonstrate the behavior formulated in Proposition~\ref{prop:1}: static analysis does not detect false negatives (Hypothesis~H1).
We further demonstrated that the commercial competitor only detects a specific subset of errors.

Our experiments with \opsc\ in degraded mode showcase the practical benefits of static code analysis. We detected true positives, some of them serious issues, in functional real-world software. At the same time, we detected no false positives, so developers should not worry about being overwhelmed with messages~(H2.1).

Out-of-the-box, a good share of getters/setters can be checked when analyzing existing software, and this share can be further increased by adding manual annotations.
We argued that we perceived the annotation effort as reasonable, and that there is even potential for automated support.  Given that the reach of \opsc\ can be considerably increased via annotations, we conclude that the effort is worthwhile in any case~(H2.2). If software is written from scratch, or during refactoring, developers can even take care to modularize their code with \opsc\ in mind.

For practical usability, the resource overhead during compilation must be reasonable. We can confirm that it is, esp.\ compared to the baseline cost imposed by the Value Checker, which is already used in professional software development. At the very least, \opsc\ is affordable enough to be integrated into offline analysis, like continuous builds or nightly builds (H2.3).

\section{Conclusion and Outlook}

As our experiments confirm, our \opsc\ checker can indeed identify real problems in database access code, even in functional real-world software.
The approach presented in this paper builds on the versatile \emph{Checker Framework}.
 Exploiting the power of this framework for extended Java type systems allows us to incorporate guarantees that other type extensions in the framework already make. For SQL query statements, this includes the SQL Quotes Checker (outlined in Section~\ref{sec:prelims_checker}) to detect SQL injection vulnerabilities.  
By systematically integrating existing checkers with database schema constraints, we may unlock entirely new opportunities in the static analysis of JDBC application code. In the following, we outline several promising avenues for future research.

\emph{Nulls.}
A promising direction is to address the gap between Java's \texttt{null} and SQL's \texttt{NULL}. Java \texttt{NullPointerExceptions} are undesirable, and while most IDEs provide nullability warnings for uninitialized variables, these checks are often very basic. The Nullness Checker~\cite{cfUses} goes beyond such simplistic analyses: it is aware of control flow, Java generics, and can suppress warnings when they are provably unnecessary. By also considering database schema constraints, we may statically verify, for example, that a JDBC setter call does not assign a \texttt{null} value to a non-nullable database column, or that a getter-call is guaranteed to return a non-\texttt{null} value.
Standardization efforts like JSpecify\footnote{\url{https://jspecify.dev/}} further highlight the importance of safe null handling in Java.

\emph{String lengths.}
The Index Checker~\cite{KelloggEA2018} tracks the length of Java strings. By comparing against length constraints derived from the database schema (e.g., the SQL type \cd{CHAR(24)}, line~17 in Example~\ref{ex:control-flow}), we can statically prevent string truncations in setter calls. By propagating such constraints, we may even prevent truncations when strings fetched from the database are displaced in the GUI. This is a common problem when user dialogs are internationalized: Developers in Europe may not notice when a Chinese character is lost, but affected end users will.

\emph{Result cardinality.}
A further opportunity is to recognize singleton queries, i.e., queries that return exactly one tuple (e.g.\ due to aggregation). Such result sets can be safely processed outside of loops. Conversely, if a query may return multiple tuples but only the first is ever processed by the Java code, developers may be advised to extend the SQL query with \texttt{LIMIT}. These queries may then be evaluated more efficiently at the side of the database.

\emph{Prepared statements.}
As outlined in Section~\ref{sec:ann-based-type-chk}, we can extend static analysis in \opsc\ to check whether all parameters in a prepared statement are indeed set. This goes beyond what current tools detect and would help identify one-off errors, especially when parameter indices are bound sequentially (see Example~\ref{ex:increment}).

\emph{Extending non-local checks.}
One limitation of our approach is that non-local accesses cannot always be checked.
Here, we see potential to further extend the reach of \opsc{} by manually annotating projects
with the \texttt{@CreatesSqlStatement} and \texttt{@Re\-trieves\-SqlResultSet} method annotations.
This could allow support for custom wrapper methods that create statements by calling JDBC methods, as shown in Example~\ref{ex:limitation}.
Extending constraint-based, whole-program type inference~\cite{cfPUnit} to the String-based \cd{@Sql} annotations would further simplify the annotation effort.

\smallskip 
Overall, we are confident that extending \opsc\ with  features as outlined above constitutes valuable contributions, not only for JDBC novices who are prone to make rookie mistakes, but also for experienced developers in charge of maintaining JDBC legacy code, as well as the growing number of ``vibe coders'' challenged to sign off on AI-generated database application code.

\bibliographystyle{ACM-Reference-Format}
\bibliography{bibliography}

\end{document}